\documentclass[manuscript=article]{achemso}
\usepackage{graphicx}% Include figure files
\usepackage{dcolumn}% Align table columns on decimal point
\usepackage{bm}% bold math
\usepackage{multirow}
\usepackage{amsmath}
\usepackage{amssymb}
\usepackage{breqn}
\usepackage[english]{babel}
\newcommand{\cmmnt}[1]{}
\def \Evec {\mathbf{E}}
\def \rvec {\mathbf{r}}
\def \nrt {n(\rvec,t)}
\def \nprt {n'(\rvec',t')}
\def \Irt {I(\rvec,t)}

\def \nsqrt {n^2(\rvec,t)}
\def \npsqrt {n'^2(\rvec',t')}
\def \Fdec {F_{decay}(\rvec',\eta_0)}

\author{T. V. Raziman}
\author{C. Peter Visser}
\affiliation{Department of Applied Physics and Institute for Photonic Integration, Eindhoven University of Technology, Eindhoven, The Netherlands}
\author{Shaojun Wang}
\affiliation{Department of Applied Physics and Institute for Photonic Integration, Eindhoven University of Technology, Eindhoven, The Netherlands}
\alsoaffiliation{MOE Key Laboratory of Modern Optical Technologies and Jiangsu Key Laboratory of Advanced Optical Manufacturing Technologies, School of Optoelectronic Science and Engineering, Soochow University, Suzhou 215006, China}
\author{Jaime G\'omez Rivas}
\author{Alberto G. Curto}
\affiliation{Department of Applied Physics and Institute for Photonic Integration, Eindhoven University of Technology, Eindhoven, The Netherlands}
\email{A.G.Curto@TUe.nl, Phone:+31-40-247-4205}
\title{Exciton diffusion and annihilation in nanophotonic Purcell landscapes}


\begin{document}
%\pagebreak

\begin{abstract}

Excitons spread through diffusion and interact through exciton-exciton annihilation.
Nanophotonics can counteract the resulting decrease in light emission. However, conventional enhancement treats emitters as immobile and non-interacting.
It neglects exciton redistribution between regions with different enhancements and the increase in non-radiative decay at high exciton densities.
Here, we go beyond the localized Purcell effect to exploit exciton dynamics and turn their typically detrimental impact into additional emission.
As interacting excitons diffuse through optical hotspots, the balance of excitonic and nanophotonic properties leads to either enhanced or suppressed photoluminescence. 
We identify the dominant enhancement mechanisms in the limits of high and low diffusion and annihilation.
Diffusion lifts the requirement of spatial overlap between excitation and emission enhancements, which we harness to maximize emission from highly diffusive excitons.
In the presence of annihilation, we predict improved enhancement at increasing powers in nanophotonic systems dominated by emission enhancement.
Our guidelines are relevant for efficient and high-power light-emitting diodes and lasers tailored to the rich dynamics of excitonic materials such as monolayer semiconductors, perovskites, or organic crystals.

\end{abstract}

%\maketitle
\section{Introduction}

Nanophotonics can improve light emission by enhancing excitation and radiative rates, and beaming  radiation~\cite{Bidault_2019,Tiguntseva_2018,Rutckaia_2017,Bucher_2019,Raziman_2019,Murai_2020,Cihan_2018, Farahani_2005,Anger_2006,Kuhn_2006, Vecchi_2009, Curto_2010, Huang_2012}.
In the conventional Purcell effect, emitters such as molecules and quantum dots are treated as localized point dipoles~\cite{Mohammadi_2008,Novotny_2012}.
Total enhancement thus benefits from the product of excitation and emission at a point, which  guides the design of nanoresonators and metamaterials made of metals and dielectrics.

For excitonic emitters, however, the picture of emission arising from non-interacting dipoles at fixed positions is incomplete.
In a variety of semiconductors, excitons are mobile and spread to large diffusion lengths compared to nanophotonic scales (10 -- 500 nm). Examples include perovskites (diffusion length $L_D \sim$ 100 -- 1000 nm)~\cite{Stranks_2013,Yao_2019,Deng_2020}, monolayer transition metal dichalcogenides (hundreds of nm)~\cite{Kumar_2014,Yuan_2017}, quantum dots (tens of nm)~\cite{Akselrod_2014,Lee_2015}, organic crystals (1 -- 100 nm for singlet excitons, 10 -- 1000 nm for triplets)~\cite{Mikhnenko_2015}, and carbon nanotubes (hundreds of nm)~\cite{Cognet_2007,Amori_2018}.
As a result, excitons can emit far from the intense near field where they originate, affecting their radiative rate.
Additionally, diffusion deteriorates emission as emitters approach defects and boundaries, where they might decay non-radiatively~\cite{Hertel_2010,Li_2018,Snaider_2018,Goodman_2020}.
A photonic modification of the radiative decay rate could decrease the effective diffusion length, thus improving performance.

Another important aspect of exciton dynamics is exciton-exciton annihilation~\cite{Yuan_2015,Amori_2018,Sun_2014}. At high exciton densities, this nonlinear process contributes to and even dominates non-radiative losses, degrading the performance of light-emitting devices at high powers~\cite{Mouri_2014,Yu_2016} and potentially preventing lasing. Annihilation thus curtails the advantages of nanophotonic intensity enhancement as well by simultaneously increasing non-radiative decay.

\begin{figure}[t]
    \centering
    \includegraphics[width=\columnwidth]{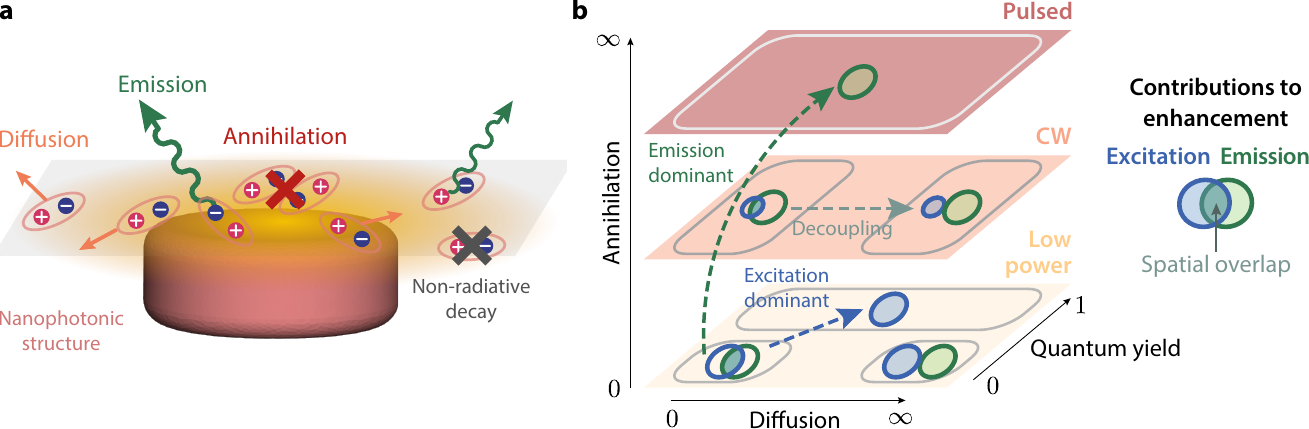}
    \caption{Impact of exciton dynamics on nanophotonic emission enhancement. (a) Excitation and emission of a thin excitonic film (gray) above an array of nanodisks in the presence of diffusion and annihilation. Excitation enhancement (orange cloud)  generates excitons that diffuse (orange arrows), annihilate (red), or decay non-radiatively (black) before radiative decay (green). (b) Contributions to the total photoluminescence enhancement of excitation, emission, and their spatial overlap (see Table~\ref{limits}). Limiting cases of diffusion, annihilation, and quantum yield with qualitatively similar enhancement behaviors are grouped into common areas. At low power, diffusion lifts the requirement of spatial overlap between excitation and emission enhancements for emitters with low quantum yield, whereas enhancement with high quantum yield is independent of diffusion or emission enhancement. At high power, the benefit of excitation enhancement decreases and even vanishes for continuous-wave and pulsed illuminations, with emission dominating the total enhancement. For continuous-wave illumination, enhancement depends on diffusion but is independent of quantum yield; for pulsed illumination, enhancement does not depend on quantum yield or diffusion.}
    \label{introduction}
\end{figure}

Here, we analyze the interplay of exciton dynamics and nanophotonic enhancement for thin films of excitonic emitters in nanostructured landscapes. We provide analytical results for enhancement under limiting cases of exciton dynamics. We demonstrate that, although diffusion and annihilation typically impede nanophotonic enhancement, it is possible to design nanostructures to overcome the lost efficiency.
Diffusion can increase photoluminescence by taking excitons to highly radiative locations when excitation and emission are spatially decoupled. 
Radiative rate enhancement can ameliorate the loss of efficiency arising from annihilation, while the interplay between annihilation and diffusion can improve performance by redistributing the local exciton density.
In summary, a careful balance of the relative strengths and spatial overlap of excitation and emission enhancements is the key to efficient excitonic-nanophotonic systems.
Our guidelines for tailoring nanophotonic structures to diffusing and annihilating excitons will aid the design of efficient light-emitting devices.

\section{Results}
\subsection{Exciton dynamics and nanophotonic enhancement}

Excitons evolve in nanophotonic environments under the combined influence of incident intensity, radiative enhancement, non-radiative decay, diffusion, and annihilation (Figure~\ref{introduction}a).
We consider excitonic emitters in ultrathin films on top of nanodisk arrays with a negligible variation of electromagnetic fields across the film thickness. The two-dimensional exciton density evolves according to the exciton dynamics equation~\cite{Yuan_2017}:

%\begin{widetext}
%\begin{equation}
%    \frac{\partial \nrt}{\partial t} =  \Irt \sigma - \left[ \Gamma_r (\rvec) + \Gamma_{nr} \right] \nrt + D \nabla^2 \nrt - \gamma \nsqrt \,,
%    \label{excitondynamics}
%\end{equation}
%\end{widetext}
\begin{dmath}
    \frac{\partial \nrt}{\partial t} =  \Irt \sigma - \left[ \Gamma_r (\rvec) + \Gamma_{nr,0} \right] \nrt \nonumber 
    + D \nabla^2 \nrt - \gamma \nsqrt \,,
    \label{excitondynamics}
\end{dmath}
where $\nrt$ is the exciton density at point $\rvec$ at time $t$, $I$ is the nanophotonically enhanced local excitation intensity at $\rvec$, $\sigma$ is the absorption coefficient, $\Gamma_r(\rvec)$ is the spatially varying radiative decay rate, $\Gamma_{nr,0}$ is the intrinsic non-radiative decay rate, $D$ is the diffusion constant, and $\gamma$ is the annihilation constant. In the absence of nanophotonic structures, the intrinsic decay rates are $\Gamma_{r,0} = \eta_0 \Gamma_0$ and $\Gamma_{nr,0} = (1-\eta_0)\Gamma_0$, where $\Gamma_0$ is the total decay rate and $\eta_0$ is the intrinsic quantum yield. The exciton decay time is $\tau_0 = 1/\Gamma_0$ and the diffusion length is $L_D = \sqrt{D\tau_0}$. We assume that the nanostructures do not modify the non-radiative decay rate $\Gamma_{nr,0}$ and the diffusion constant $D$, although our model can incorporate such changes as well. We also neglect saturation of absorption at high power.

To compare systems with different excitonic and nanophotonic properties and extract the universal behavior of nanophotonic systems in the presence of exciton dynamics, we non\nobreakdash-dimensionalize Equation~(\ref{excitondynamics}). We identify physically relevant scales of exciton density, incident power, length, and time in the system to scale the variables $n$, $I$, $\rvec$, and $t$ with these values:
\begin{itemize}
    \item $n' = n/n_0$, where $n_0 = \Gamma_0/\gamma$ is the exciton density at which $\Gamma_0 n$ (the intrinsic total decay, which includes radiative and non-radiative rates but not annihilation) equals $\gamma n^2$ (the density-dependent annihilation),
    \item $I' = I/I_0$, where $I_0 = \Gamma_0n_0/\sigma$ is the incident continuous-wave power at which $I \sigma$ (the exciton generation) equals $\Gamma_0 n_0$ (the intrinsic decay at $n=n_0$),
    \item $\rvec' = \rvec/P$, where $P$ is the period of the nanophotonic structures,
    \item $t' = t/\tau_0$, where $\tau_0$ is the exciton decay time.
\end{itemize}
The characteristic scales depend on both excitonic and nanophotonic properties. Note that we perform scaling with respect to intrinsic properties (in the absence of nanostructures), except the length scale, which is the period of the array. Expressing Equation~(\ref{excitondynamics}) in terms of the primed variables, we obtain the non-dimensionalized exciton dynamics equation

\begin{dmath}
    \frac{\partial \nprt}{\partial t'} =  F_{ex}(\rvec') I'(t') - \Fdec \nprt \nonumber
    + D' \nabla^{\prime 2} \nprt 
    - \npsqrt \,,
    \label{nondimensional}
\end{dmath}
where $F_{ex}$ and $F_{em}$ are the local nanophotonic excitation and radiative rate enhancements, and $F_{decay} = \eta_0 F_{em} + 1 - \eta_0$ is the localized total decay rate enhancement.
The non-dimensionalized diffusion constant $D'=D \tau_0/P^2$ is related to the diffusion length and the period of the nanophotonic structures by $D' = \left(L_D/P\right)^2$. The annihilation rate $\gamma$ does not appear explicitly and is part of the characteristic incident power $I'=I\gamma \sigma/\Gamma_0^2$, demonstrating the utility of non-dimensionalization in comparing different systems. Excitonic materials that differ only in the annihilation rate $\gamma$ will behave identically in a nanophotonic system except for scaling of incident power. We perform electromagnetic simulations using the surface integral equation (SIE) method and solve the non-dimensionalized exciton dynamics equation (\ref{nondimensional}) to demonstrate the impact of exciton diffusion and annihilation on nanophotonic photoluminescence enhancement.
We illustrate the diverse behavior and the possible scenarios of exciton dynamics using a variety of nanostructures.

\begin{figure}[t]
    \centering
    \includegraphics[width=0.65\columnwidth]{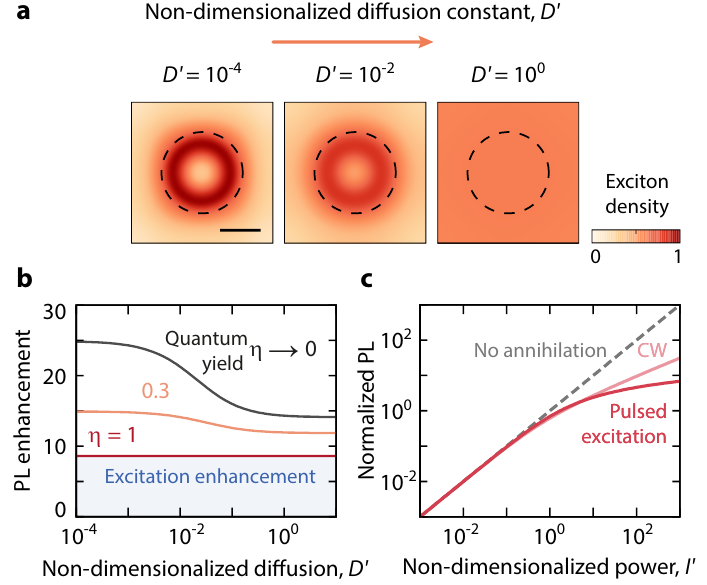}
    \caption{Exciton diffusion and annihilation usually lead to reduced photoluminescence. (a) Exciton density of emitters with intrinsic quantum yield $\eta_0$ = 1 above an array of silicon nanodisks with radius $R$~=~100, height $H$~=~75, and period $P$~=~365~nm. A low-power, continuous plane wave is incident from below the disks at $\lambda$~=~553~nm. Scale bar 100 nm. (b) The nanophotonic photoluminescence enhancement of the nanodisk array decreases due to diffusion. The blue area indicates enhancement due to excitation only. (c) In the absence of nanophotonic structures, photoluminescence scales sublinearly with incident power due to annihilation.}
    \label{typical}
\end{figure}

First, we analyze how diffusion affects it by spreading excitons out.
We study excitonic emitters consisting of orientationally averaged dipoles above an array of silicon nanodisks.
Although exciton density and diffusion are two-dimensional because the exciton film is thin, the exciton dipole moment can have components out of the plane -- therefore, we average the orientations in all three dimensions~\cite{Raziman_2016}.
We assume perfect collection efficiency of the emission.
We illuminate with a continuous-wave source at low power so that annihilation is initially negligible.

For low diffusion constants, the exciton density concentrates near the edge of the nanodisk, corresponding to the local excitation profile of an electric dipole in the nanodisk (Figure~\ref{typical}a).
As the non-dimensionalized diffusion constant $D'$ increases, the exciton density distribution expands, eventually becoming uniform over the unit cell.
In this case, diffusion suppresses the total photoluminescence enhancement by taking excitons from regions of high radiative enhancement to positions of low enhancement (Figure~\ref{typical}b).
The total photoluminescence is a combination of excitation and emission enhancements. Immobile emitters with low intrinsic quantum yield benefit from both excitation and emission enhancements, whereas only the increased excitation is relevant for emitters with high quantum yield~\cite{Sun_2009, Kinkhabwala_2009}. The Purcell effect can be strongly modified upon diffusion, while the excitation is unaffected by diffusion in the absence of saturation effects. The impact of diffusion is the strongest for emitters with low quantum yield because of their larger contribution from radiative enhancement. For emitters with high intrinsic quantum yield, the Purcell effect enhances only the decay rate and not the photoluminescence. Diffusion reduces this enhancement (Supporting Figure~S1a).
%Fig. S1a

In the absence of nanostructures, exciton-exciton annihilation suppresses photoluminescence by opening an additional non-radiative channel at high excitation powers and exciton densities (Figure~\ref{typical}c). Compared to continuous-wave excitation, pulsed excitation creates higher instantaneous exciton densities, thereby reducing emission even further. Nanophotonic structures can ameliorate this deterioration of emission, as we shall discuss later. Additionally, the quick initial decay due to the high exciton density shortens the decay time considerably (Supporting Figure~S1b).
%Fig. S1b

\subsection{Enhancing emission through diffusion}

\begin{table}[t]
    \centering
    \bgroup
    \caption{\label{limits}Nanophotonic photoluminescence enhancement ($F_{tot}$) in the presence of exciton dynamics: limiting cases of diffusion, quantum yield, and incident power. $\left<x\right>$ represents the spatial average of the quantity $x$ in the unit cell. Spatial averages of products of excitation and emission enhancements such as $\left<F_{ex} \cdot F_{em}\right>$ indicate the benefit of their spatial overlap. Diffusion decouples the enhancements spatially, turning the expressions into products of spatial averages such as $\left<F_{ex}\right> \left<F_{em}\right>$.}
    \def\arraystretch{1.25} 
    \begin{tabular}{|c|c|c|c|c|}
%    \multicolumn{5}{c}{\textbf{Photoluminescence enhancement} $F_{tot}$} \\
    \hline
    \multirow{3}{*}{\textbf{Diffusion}} & \multirow{3}{*}{\begin{tabular}{c}
    \textbf{Quantum} \\ \textbf{yield}
    \end{tabular}} & \multicolumn{3}{c|}{\textbf{Incident power (with annihilation)} } \\
    \cline{3-5}
    & & \multirow{2}{*}{$I' \rightarrow 0$} & \multicolumn{2}{c|}{$I' \rightarrow \infty$} \\
    \cline{4-5}
    \hphantom{Continuous} & \hphantom{Continuous} & \hphantom{Continuous} & Continuous & Pulsed \\
    \hline
    \multirow{2}{*}{$D' = 0$} & $\eta_0 \rightarrow 0$ & $\left\langle F_{ex} \cdot F_{em} \right\rangle$ &  & \hphantom{Continuous} \\
    \cline{2-3}
    & $\eta_0 = 1$ &  $\left\langle F_{ex} \right\rangle$ &  \multirow{-2}{*}{$\left\langle \sqrt{F_{ex}} \cdot F_{em} \right\rangle$}  &  \\
    \cline{1-4}
    \multirow{2}{*}{$D' \rightarrow \infty$} & $\eta_0 \rightarrow 0$ &  $\left\langle F_{ex} \right\rangle \left\langle F_{em} \right\rangle$ &  &  \\
    \cline{2-3}
    & $\eta_0 = 1$ &  $\left\langle F_{ex} \right\rangle$ &  \multirow{-2}{*}{$\left\langle \sqrt{F_{ex}} \right\rangle \left\langle F_{em} \right\rangle$}  & \multirow{-4}{*}{$F_{tot} = \left\langle F_{em} \right\rangle$ } \\
    \hline
    \end{tabular}
    \egroup
\end{table}

\begin{figure}[t]
    \centering
    \includegraphics[width=0.6\columnwidth]{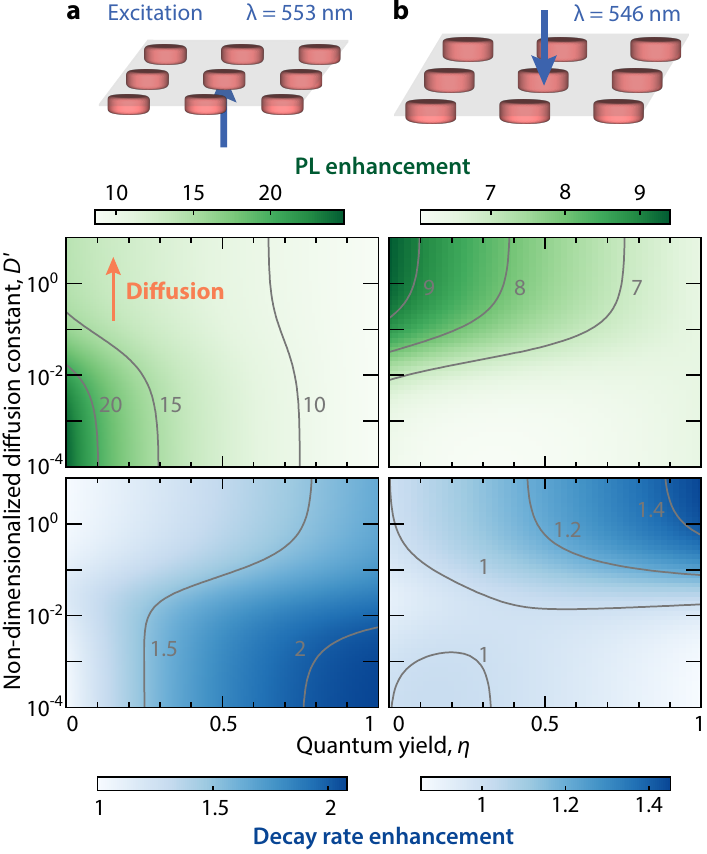}
    \caption{Diffusion can both improve or degrade photoluminescence depending on the nanophotonic system. Photoluminescence and total decay rate enhancements under low-power, pulsed illumination of excitons above arrays of silicon nanodisks: (a) illuminated from below with $R$~=~100, $H$~=~75, $P$~=~365~nm; (b) illuminated from above with $R$~=~120, $H$~=~90, $P$~=~445~nm. In gray, contours of constant enhancement with their respective values. Excitation and emission wavelengths are equal.}
    \label{diffusion}
\end{figure}

Generally, the deterioration of nanophotonic enhancement with diffusion is due to losing the advantage of spatial overlap between excitation and emission enhancements.
To understand the contribution of each process and their overlap, we solve the non-dimensionalized exciton dynamics equation (\ref{nondimensional}) analytically for limiting cases of quantum yield, diffusion, and annihilation (Supporting Section S1). We list the photoluminescence enhancement for these extremes in Table~\ref{limits} and depict the contributions from excitation and emission and their overlap in Figure~\ref{introduction}b, where the levels indicate the increasing role of annihilation.
When the incident power is much lower than $I_0=\Gamma_0n_0/\sigma$, annihilation is negligible compared to intrinsic decay (bottom level in Figure~\ref{introduction}b). In this regime of negligible annihilation ($I' \rightarrow 0$) and diffusion length much smaller than the period ($D' \rightarrow 0$), the total enhancement for emitters with poor efficiency ($\eta_0 \rightarrow 0$) is $\left<F_{ex} (\rvec) \cdot F_{em}(\rvec)\right>$, which is the spatial average of the product of local enhancements in the unit cell. Hence, in the absence of diffusion, we obtain high total enhancement if the excitation and emission significantly overlap.

In the regime of high diffusion, however, the total enhancement becomes $\left<F_{ex} (\rvec)\right> \left<F_{em} (\rvec)\right>$, which is the product of the average values of excitation and emission in the unit cell.
The spatial overlap of the enhancement factors is then no longer of benefit.
As a result, the photoluminescence enhancement typically worsens with diffusion for low-efficiency emitters (Figure~\ref{diffusion}a).
Although emitters with high quantum yield do not suffer a similar loss of enhancement because their emission efficiency does not change, their decay rate deteriorates with increasing diffusion (Figure~\ref{diffusion}a). Surprisingly, diffusion can modify the decay rate of emitters even in the limit of zero quantum yield (Supporting Figure~S2).
%Fig. S2

Diffusion can also improve emission by removing the spatial overlap between enhancement contributions. By controlling the excitation conditions such as the angle of incidence, polarization, or wavelength, we can lift the requirement of spatial overlap for maximum photoluminescence. As a first example, we spatially decouple excitation and emission by exploiting the angular pattern of emission in an array with a different geometry (Figure~\ref{diffusion}b and Supporting Figure~S3).
%Fig. S3
The excitation profile shows a strong front-back asymmetry because the high refractive index of silicon causes retardation of electromagnetic fields along its height~\cite{Raziman_2019}. We illuminate the nanodisks from above to benefit from this asymmetry.
The photoluminescence increases with diffusion at low $\eta_0$, and so does the decay rate enhancement at high $\eta_0$.
Nanostructures designed under the assumption of immobile emitters can thus behave differently with diffusing excitons.
Whether the impact of diffusion on enhancement is beneficial or detrimental depends on the nanophotonic system.
Nanostructures aiming at maximum output from excitonic materials should thus take diffusion into account due to their spatially dependent enhancements~\cite{Cihan_2018, Bucher_2019, Mey_2019, Sortino_2019}.

\begin{figure}[t]
    \centering
    \includegraphics[width=0.6\columnwidth]{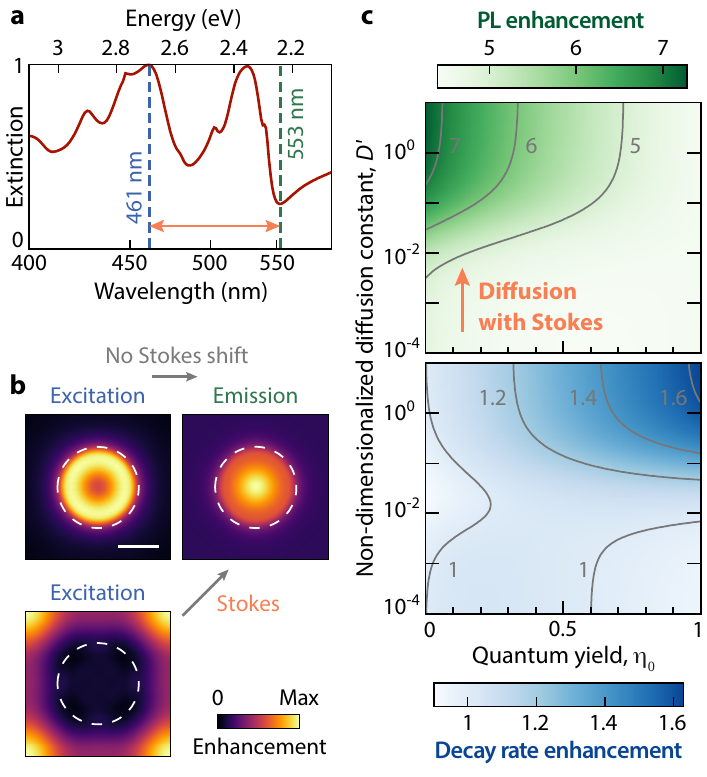}
    \caption{A Stokes shift can decouple excitation and emission enhancements at different wavelengths to improve photoluminescence through diffusion. (a) Extinction spectrum of the array of silicon nanodisks in Figure~\ref{diffusion}a with multiple resonances. (b) The maps of excitation and emission enhancements at $\lambda_{em}$~=~553~nm overlap significantly. Changing the excitation wavelength to 461~nm decouples the excitation enhancement at $\lambda_{ex}$ from the emission at $\lambda_{em}$. Scale bar 100 nm. (c) Enhancements of photoluminescence and total decay rate under excitation at $\lambda_{ex}$ and emission at $\lambda_{em}$ including the Stokes shift above.}
    \label{stokes}
\end{figure}
 
\begin{figure}[t]
    \centering
    \includegraphics[width=\textwidth]{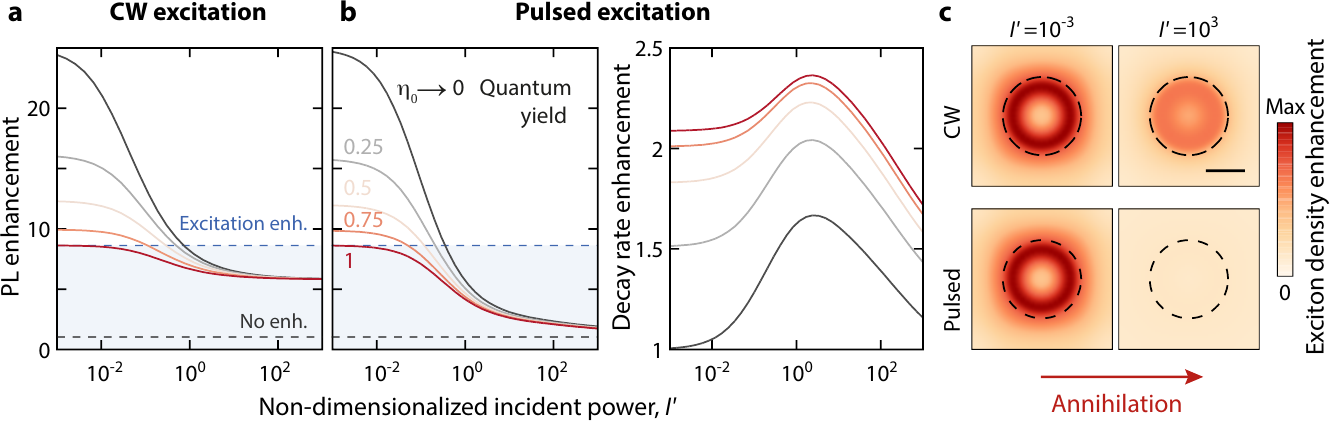}
    \caption{Exciton-exciton annihilation typically suppresses photoluminescence enhancement at increasing excitation powers. (a) Photoluminescence enhancement for the array of silicon nanodisks in Figure~\ref{diffusion}a as a function of incident power under continuous-wave illumination. Blue and gray lines denote the levels of excitation enhancement and no enhancement. (b) Enhancement of photoluminescence and total decay rate under pulsed illumination. (c) Change in exciton density enhancement averaged over time with increasing incident power.}
    \label{annihilation}
\end{figure}

As a second case of enhancement through diffusion, we exploit the Stokes shift between excitation and emission wavelengths to decouple them spatially.
We utilize the diversity of resonances supported by arrays of silicon nanodisks. (Figure~\ref{stokes}a).
For zero detuning with emission and excitation at $\lambda_{em}=\lambda_{ex}$~=~553~nm, the excitation and radiative enhancements  are both strongest above the disk (Figure~\ref{stokes}b).
The exciton density is highest near the edge because of the relatively high excitation enhancement there (Figure~\ref{typical}a).
Due to the spatial overlap between enhancements, diffusion decreases photoluminescence as it transports excitons from regions of high excitation to areas of low radiative enhancements (Figure~\ref{diffusion}a).
However, if the excitonic material has a Stokes shift between excitation at $\lambda_{ex}$ =~461 and emission at $\lambda_{em}$~=~553 nm, the excitation enhancement is almost completely decoupled from the emission enhancement (Figure~\ref{stokes}b).
Diffusion thus takes the excitons generated at $\lambda_{ex}$ to regions of high radiative enhancement at $\lambda_{em}$, improving photoluminescence and decay rate compared to immobile excitons (Figure~\ref{stokes}c).
Emitters typically have a Stokes shift between excitation and emission, giving us a handle to turn diffusion to our advantage.

\subsection{Overcoming annihilation through nanophotonic enhancement}

So far, we have only considered the effects of diffusion on nanophotonic enhancement. Next, we add exciton-exciton annihilation, which typically suppresses photoluminescence.
As the incident power of a continuous-wave source increases, photoluminescence enhancement usually decreases for all quantum yields in the absence of diffusion (Figure~\ref{annihilation}a for the array in Figure~\ref{diffusion}a).
At high power, the total nanophotonic enhancement falls even below the excitation enhancement (blue line).
Exciton-exciton annihilation increases nonlinearly with power, suppressing the effect of excitation enhancement and reducing the steady-state exciton density enhancement (Figure~\ref{annihilation}c).
At low power, the photoluminescence enhancement is due to $F_{ex}$ for high-$\eta_0$ emitters, whereas it arises from the product of $F_{ex}$ and $F_{em}$ for low-$\eta_0$ emitters. In contrast, at high power, the photoluminescence enhancement is the product of $\sqrt{F_{ex}}$ and $F_{em}$ independent of the quantum yield because exciton-exciton annihilation becomes the dominant non-radiative decay channel (Table~\ref{limits}).

The suppression of photoluminescence is even stronger for pulsed excitation (Figure~\ref{annihilation}b), where annihilation neutralizes the excitation enhancement completely as manifest in the time-averaged exciton density (Figure~\ref{annihilation}c).
The emission enhancement still results in increased photoluminescence compared to the bare emitters. Exciton density enhancement on the nanostructure modifies the total decay rate via exciton-exciton annihilation, although the effect is neutralized once again at very high incident power (Figure~\ref{annihilation}b).

\begin{figure}[t!]
    \centering
    \includegraphics[width=0.6\columnwidth]{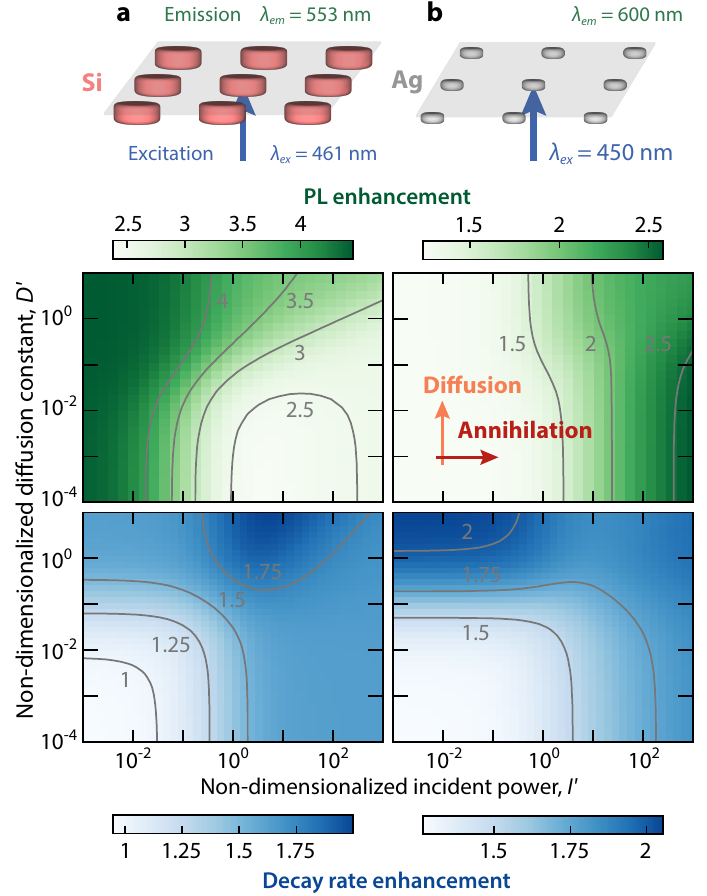}
    \caption{Interplay between nanophotonic properties, annihilation, and diffusion. Increasing incident power can either decrease or increase nanophotonic emission enhancement.  (a,b) Geometry,  photoluminescence enhancement under continuous-wave illumination, and  total decay rate enhancement under pulsed illumination, for emitters with $\eta_0 = 1$ above: (a) silicon nanodisks with $R$~=~100, $H$~=~75, $P$~=~365~nm; (b) silver nanodisks with $R$~=~50, $H$~=~40, $P$~=~350~nm.}
    \label{diffusionannihilation}
\end{figure}

Diffusion can alleviate part of the detrimental effects of annihilation by smearing the hotspots of exciton density. Although emitters with high intrinsic quantum yield have diffusion-independent photoluminescence enhancement at low incident power (Figure~\ref{stokes}c), the drop in enhancement with increasing power is much slower for highly diffusing excitons (Figure~\ref{diffusionannihilation}a).

Next, we demonstrate that it is possible to improve performance even as annihilation becomes dominant.
At low incident power, high-$\eta_0$ emitters benefit only from excitation enhancement whereas at high power, the effect of excitation diminishes and emission enhancement becomes dominant (Table~\ref{limits}). 
Therefore, nanophotonic structures with emission enhancement comparable to or higher than excitation enhancement offer improved photoluminescence enhancement with increasing power.
This counter-intuitive behavior arises from the increasing benefit of emission enhancement as a strong non-radiative decay channel opens at high exciton densities.
We exemplify such a case with an array of silver nanoparticles which has significantly higher emission enhancement compared to excitation enhancement (Supporting Figure~S4). Indeed, as the incident power increases, emitters above the array of silver nanoparticles benefit from increasing photoluminescence enhancement. (Figure~\ref{diffusionannihilation}b).
Although we have shown improved performance here for plasmonic nanoparticles, dielectric nanostructures with strong emission enhancements behave similarly (Supporting Figure~S5).
High-$\eta_0$ emitters with exciton-exciton annihilation are important for light-emitting devices, which suffer from efficiency loss at high powers. The ability to reduce the impact of annihilation on emission through the combination of nanophotonic design and diffusion is thus of practical interest.

\subsection{Nanophotonic enhancement in relevant excitonic materials}

Finally, we show that the non-dimensionalized limits of diffusion and annihilation become significant for light emission from realistic excitonic emitters. We tabulate reported excitonic parameters for representative materials and calculate their characteristic exciton density $n_0$, incident power $I_0$ and the non-dimensionalized diffusion constant $D'$ (Supporting Table~S1). For transition metal dichalcogenide monolayers and two-dimensional perovskites, $D'$ is of the order of unity. As a result, the excitons spread through the entire unit cell before they decay. Monolayer semiconductors with low quantum yield thus benefit from additional enhancement in nanostructures designed for diffusive excitons (Supporting Figure~S6). Emitters with  high quantum yield also suffer from strong annihilation . The incident power at which annihilation becomes dominant, $I_0$, is very low (of the order of nW/$\mu$m$^2$ -- $\mu$W/$\mu$m$^2$), especially for transition metal dichalcogenides (Supporting Table~S1).  As a result, they benefit from additional enhancement in nanophotonic systems designed for materials with high annihilation (Supporting Figure~S7).

Although we focused our analysis on excitons in thin films, the general principles also apply to other geometries.
Nanowires support one-dimensional diffusion and can be placed along directions of high excitation and emission enhancements to obtain stronger photoluminescence (Supporting Figure~S10).
In the case of thick excitonic materials around nanophotonic structures, exciton diffusion will be three-dimensional, and the decay of the evanescent near field away from the nanostructure plane will also play a role.
In addition to photonic enhancement, material interfaces can modify intrinsic decay, diffusion, and annihilation through doping, dielectric screening, or  phonons~\cite{Yu_2016, Hoshi_2017, Schneider_2018, Fu_2019, Rhodes_2019, Godiksen_2020,  Goodman_2020}. Our model can accommodate such effects through the use of locally varying excitonic parameters modified by the environment. It is also possible to prevent such environmental modification of excitonic parameters with a thin dielectric spacer.
At high exciton densities in transition metal dichalcogenide monolayers, exciton-phonon effects modify diffusion, resulting in halo formation~\cite{Kulig_2018,Glazov_2019,Perea-causin_2019}. Making the diffusion constant dependent on the exciton density~\cite{Kulig_2018} in the exciton dynamics equation can model such behavior.
Some materials show superdiffusive behavior where the exciton density spreads faster than the prediction from linear diffusion. A rate equation model can describe such behavior by making the diffusion constant dependent on time or exciton density~\cite{Najafi_2017,Flekkoy_2021}.
In materials such as organic semiconductors where there are multiple varieties of excitons, incorporating their interconversion into the rate equation can identify effects such as negative diffusion~\cite{Berghuis_2021}.
Additional phenomena could further exploit diffusion to improve performance.
Diffusion spreads the excitons away from the nanostructures, where Ohmic losses in the nanophotonic system are stronger. The reduction of absorption losses could therefore further contribute to total enhancement due to diffusion.
Analogous to annihilation, saturation effects at higher excitation powers can suppress the enhancement~\cite{Kern_2012} and could be similarly overcome through diffusion
On the other hand, designing nanophotonic systems with extremely high emission enhancements can counteract the effect of diffusion by reducing the lifetime and thus the diffusion length (Supporting Figure~S9).
Last, applying strain on monolayer semiconductors can result in exciton funneling, the directional transport of excitons towards regions of high strain~\cite{Castellanos-gomez_2013, Liu_2019, Moon_2019, Sortino_2019, Sortino_2020}.
Using nanostructures as sources of both strain and nanophotonic enhancement promises a new direction to control light-matter interaction.

\section{Conclusion}
We have combined exciton dynamics and nanophotonics to illustrate the range of scenarios beyond the conventional Purcell effect for photoluminescence enhancement in the presence of diffusion and annihilation.
Although usually detrimental for light emission, careful nanophotonic design can turn diffusion into an advantage and partly mitigate the detrimental effect of exciton-exciton annihilation.
We presented analytical expressions of enhancement for the limiting regimes of diffusion and annihilation to formulate the conditions for improved exciton photoluminescence.
Removing the spatial overlap between excitation and emission enhancements can improve emission from diffusive excitons in nanostructured landscapes -- for instance, by including a Stokes shift between excitation and emission wavelengths. Similar decoupling is possible through other strategies such as excitation and emission at different angles.
It is also possible to alleviate the detrimental effects of exciton-exciton annihilation by tuning the relative strengths of excitation and emission enhancements and by capitalizing on diffusion to reduce the local exciton density. The approach could be extended to electrically generated excitons, which can create highly localized exciton distributions near nanoscale contacts.
Additionally, solar cells could benefit from maximizing excitation enhancement near the nanostructures while  redistributing the exciton density through diffusion to reduce the loss of excitons through emission and annihilation.

Our results demonstrate the importance of tailoring nanophotonic structures to specific exciton dynamics for maximal performance. As several excitonic materials consist of nano- or microcrystals exhibiting nanophotonic resonances, their photonic properties also have implications for understanding and quantifying exciton dynamics in nanomaterials. The operation principles and limiting regimes that we provide for diffusion and annihilation in nanophotonic landscapes can thus guide the design of efficient and high-power devices. Our findings apply to light-emitting diodes or lasers using relevant families of semiconductors in the form of nanoparticles, nanowires, thin films, or monolayers.

\section{Methods}

\subsection{Electromagnetic simulations}
We perform the electromagnetic simulations using the surface integral equation (SIE) method for periodic nanostructures~\cite{Gallinet_2010,Raziman_2015}. We use the permittivities of silicon and silver from Green~\cite{Green_2008} and Johnson and Christy~\cite{Johnson_1972}. We set a homogeneous relative permittivity $\epsilon_r=1.5$ for the background medium as the geometric mean of air and glass to approximate the effect of a substrate. We apply a realistic rounding radius of 20 nm to the sharp edges of the nanodisks.

We treat the emitters as electric dipole sources lying on a plane 5 nm above the nanodisks. To compute the excitation enhancement $F_{ex}$, we illuminate the system with a plane wave under normal incidence from above or below depending on optimal excitation conditions and evaluate the electric field $\Evec$ on the plane above the nanostructure. The excitation enhancement for dipolar emitters orientationally averaged in three dimensions is $|\Evec|^2/|\Evec_0|^2$, where $\Evec_0$ is the electric field in the absence of the nanostructures.

The emission enhancement $F_{em}$ is the integral of the power radiated in all directions $(\theta,\phi$), normalized to the same quantity in the absence of nanostructures. We compute the dipole radiation in a given direction $(\theta,\phi)$ using electromagnetic reciprocity~\cite{Chew_1990} by evaluating the field intensity at the location of the emitter under illumination by a plane wave incident from the same direction~\cite{Anttu_2016, Kivisaari_2018, Vaskin_2018}. This method assumes no absorption losses in the nanodisks that might reduce antenna radiation efficiency. The emission enhancement of a dipole depends on its orientation. We compute the average emission enhancements for emitters along all possible orientations in three dimensions, integrating total photoluminescence in all directions. 

For emitters such as transition metal dichalcogenides where the dipoles are oriented in the plane, excitation enhancement requires using the in-plane projection of the electric field. Additionally, orientational averaging should then be performed in two dimensions. With these modifications, our treatment applies to two-dimensional excitonic materials as well and does not change the results qualitatively (Supporting Figure~S8).

\subsection{Numerical solution of exciton dynamics}
To solve the non-dimensionalized exciton dynamics equation (\ref{nondimensional}) numerically, we discretize the exciton density into a grid with non-dimensionalized coordinates: $n'(x_i', y_j')$ where $x_i' = i/2N, y_j' = j/2N$ for $(i,j) \in \{-N, \ldots N\}$.
We choose the value of $N$ in each simulation to obtain $5$~nm spatial resolution. As a result of periodicity, the exciton densities are equal at opposite edges of the unit cell (indices $-N$ and $N$). In the limiting case of low incident power, the quadratic annihilation term vanishes, and we obtain a linear differential equation in $n'$. Under continuous-wave illumination, in the steady state, we have $\left[F_{decay} - D' \nabla^{\prime 2} \right] n' = F_{ex} I'\,,$
where we have dropped the explicit spatial dependence. Taking the spatial Fourier transform,

\begin{equation}
    \left[\widetilde{F}_{decay}\circledast + D' q^{\prime 2} \right] \widetilde{n}' = \widetilde{F}_{ex} I'
    \label{conveq}
\end{equation}
where the quantities with a tilde (such as $\widetilde{n'}$) are Fourier transforms of the real-space quantities, $\widetilde{F}_{decay}\circledast$ is the circular convolution matrix for $\widetilde{F}_{decay}$, and the matrix $q^{\prime 2}$ is the squared momentum in the Fourier transform of the discrete Laplace operator, with elements $q_{l,m}^{\prime 2}= 4\pi^2(l^2+m^2)$. Inverting the matrix on the left-hand side of Equation~(\ref{conveq}) gives the steady-state exciton density. Its eigenvalues describe the time evolution under pulsed illumination, providing the total decay rate enhancement.

In the presence of annihilation, we can no longer use linear methods. Hence we let the system evolve explicitly according to Equation~(\ref{nondimensional}) using the forward Euler method until $T=10 \tau_0$. Under continuous-wave illumination, the system reaches a steady state by this time. We model pulsed excitation using an ultrashort impulse $I' \delta(t')$ so that the exciton density instantaneously becomes $n'(\rvec',0) = F_{ex}(\rvec') I'$. We ensure that the high initial decay rates do not result in numerical errors by using an adaptive time step that limits the maximum relative change in exciton density at a location within a time step to one percent. The total photoluminescence from the unit cell is then

\begin{equation}
    \mathrm{PL} = \eta_0 \iiint n'(\rvec',t') F_{em}(\rvec') dt' d\rvec^{\prime 2} \,.
\end{equation}
We can also calculate the decay time as the mean lifetime of emission from the temporal decay of photoluminescence:

\begin{equation}
    \tau = \frac{\eta_0 \iiint n'(\rvec',t') F_{em}(\rvec') t' dt' d\rvec^{\prime 2}}{\mathrm{PL}} \,.
\end{equation}
The total decay rate enhancement is the ratio of the decay time without the nanostructure to the average decay time in the presence of the nanostructure.

\medskip
\noindent \textbf{Supporting Information} \par %Please delete the Suppporting Information statement if it is not applicable. Please supply Supporting Information in another file. Supporting information should not be provided in .tex format
\noindent Supporting Information is available from the Wiley Online Library or from the author.

% Acknowledgements
\medskip
\noindent \textbf{Acknowledgements} \par %delete if not applicable))
\noindent We thank Rasmus H. Godiksen for illuminating discussions. This work was financially supported by the Netherlands Organization for Scientific Research (NWO) through Gravitation grant ``Research Centre for Integrated Nanophotonics'' (024.002.033), START-UP grant (740.018.009), and  Innovational Research Incentives Scheme (VICI Grant nr. 680-47-628). S. Wang was supported by
Priority Academic Program Development (PAPD) of Jiangsu Higher Education Institutions. Simulations in this work were carried out on the Dutch national e-infrastructure with the support of SURF Cooperative. 

\section*{Conflict of interest}
The authors declare that they have no conflict of interest.

\bibliography{references}

\end{document}

\maketitle
\section{Analytical solution of the non-dimensionalized exciton dynamics equation for limiting cases}

In the low incident power limit $I' \rightarrow 0$, both continuous and pulsed excitation result in equal enhancement factors because the superposition principle applies due to linearity. Under continuous illumination and in the absence of diffusion ($D'=0$), we have the exciton density in the steady state
\begin{equation*}
    n'(\rvec'; I' \rightarrow 0, D' = 0) = I' \frac{F_{ex}(\rvec')}{\Fdec}\,.
\end{equation*}
The photoluminescence is then
\begin{equation*}
    \mathrm{PL}(\rvec'; I' \rightarrow 0, D' = 0) = \eta_0 I' \frac{F_{ex}(\rvec')F_{em}(\rvec') }{\Fdec}\,.
\end{equation*}
By spatially averaging the photoluminescence in the unit cell and normalizing by its value in the absence of enhancements, we obtain the photoluminescence enhancement
\begin{equation}
    F_{tot}(I' \rightarrow 0, D' = 0) = \left< \frac{F_{ex}(\rvec')F_{em}(\rvec')}{\Fdec} \right> \,.
\end{equation}
Under the limiting cases of quantum yield ($\eta_0 \rightarrow 0$) and $\eta_0 = 1$, this expression reduces to $\left< F_{ex}(\rvec')F_{em}(\rvec')\right>$ and $\left< F_{ex}(\rvec')\right>$, respectively.

In the infinite diffusion limit ($D' \rightarrow \infty$), the $F_{ex}(\rvec') I'$ excitons immediately migrate through the unit cell making the exciton density uniform. Similarly, diffusion smoothens out any variations due to unequal decay rates. Hence, the exciton density becomes
\begin{equation*}
    n'(\rvec'; I' \rightarrow 0, D' \rightarrow \infty) = I'\frac{\left<F_{ex}(\rvec')\right> }{\left<\Fdec\right>}\,,
\end{equation*}
giving the total photoluminescence enhancement
\begin{equation}
    F_{tot}(I' \rightarrow 0, D' \rightarrow \infty) = \frac{\left<F_{ex}(\rvec')\right>\left<F_{em}(\rvec')\right>}{\left<\Fdec\right>} \,.
\end{equation}
Under the limiting cases of quantum yield $\eta_0 \rightarrow 0$ and $\eta_0 = 1$ , we obtain total enhancements $\left< F_{ex}(\rvec')\right> \left<F_{em}(\rvec')\right>$ and $\left< F_{ex}(\rvec')\right>$, respectively.

At high incident powers for which annihilation becomes significant, we need to retain the quadratic term. Under continuous-wave illumination and in the absence of diffusion, in the steady state we have 
\begin{equation*}
    n^{\prime 2}(\rvec'; D'=0 ;\mathrm{CW}) + \Fdec n'(\rvec') - F_{ex}(\rvec') I' = 0 \,,
\end{equation*}
from which we can solve for $\nprt$
\begin{align*}
    n'(\rvec'; D'=0 ;\mathrm{CW}) 
    =\frac{ \displaystyle\sqrt{\Fdec^2 + 4 I' F_{ex}(\rvec')} - \Fdec}{2} \,.
\end{align*}
In the limiting case of very high incident power ($I' \rightarrow \infty$), where exciton-exciton annihilation dominates emission and non-radiative decay, we have
\begin{equation*}
    n'(\rvec'; I' \rightarrow \infty, D'=0 ;\mathrm{CW}) = \sqrt{I' F_{ex}(\rvec')} \,,
\end{equation*}
giving a total photoluminescence enhancement independent of quantum yield
\begin{equation}
    F_{tot}(I' \rightarrow \infty, D'=0 ;\mathrm{CW}) = \left<\sqrt{F_{ex}(\rvec')} F_{em}(\rvec)\right> \,.
\end{equation} The limit of infinite diffusion ($D' \rightarrow \infty$) follows from similar arguments as before,
\begin{equation}
    F_{tot}(I' \rightarrow \infty, D' \rightarrow \infty ;\mathrm{CW}) = \left<\sqrt{F_{ex}(\rvec')} \middle> \middle< F_{em}(\rvec)\right> \,.
\end{equation}

Under pulsed illumination, we have an initial exciton density $n'(\rvec',0) = F_{ex}(\rvec') I'$. Solving the non-dimensionalized exciton dynamics equation analytically for this initial condition under zero diffusion yields~\cite{Shaw_2008}
\begin{align*}
    n'(\rvec',t'; I' \rightarrow \infty, D'=0 ;\mathrm{pulsed})
    =\frac{\Fdec F_{ex}(\rvec')I' \exp[-\Fdec t']}{\Fdec + F_{ex}(\rvec') I'\left\{1 -\exp[-\Fdec t']\right\}} \,.
\end{align*}
Multiplying the exciton density with the radiative decay rate and integrating it over time gives the total photoluminescence from the pulse,
\begin{dmath*}
    \mathrm{PL}(\rvec'; I' \rightarrow \infty, D'=0 ;\mathrm{pulsed}) = 
    F_{em}(\rvec') \eta_0 \log\left[1 + \frac{F_{ex}(\rvec')I'}{\Fdec} \right] \,.
\end{dmath*}
The total photoluminescence enhancement in the limit $I' \rightarrow \infty$ is then
\begin{equation}
    F_{tot}(I' \rightarrow \infty, D' = 0; \mathrm{pulsed}) = \left< F_{em}(\rvec') \right> \,.
\end{equation}
The same enhancement occurs for infinite diffusion.

\section{Modification of decay rates by diffusion and annihilation}

\begin{figure}[t]
    \centering
    \includegraphics[width=\textwidth]{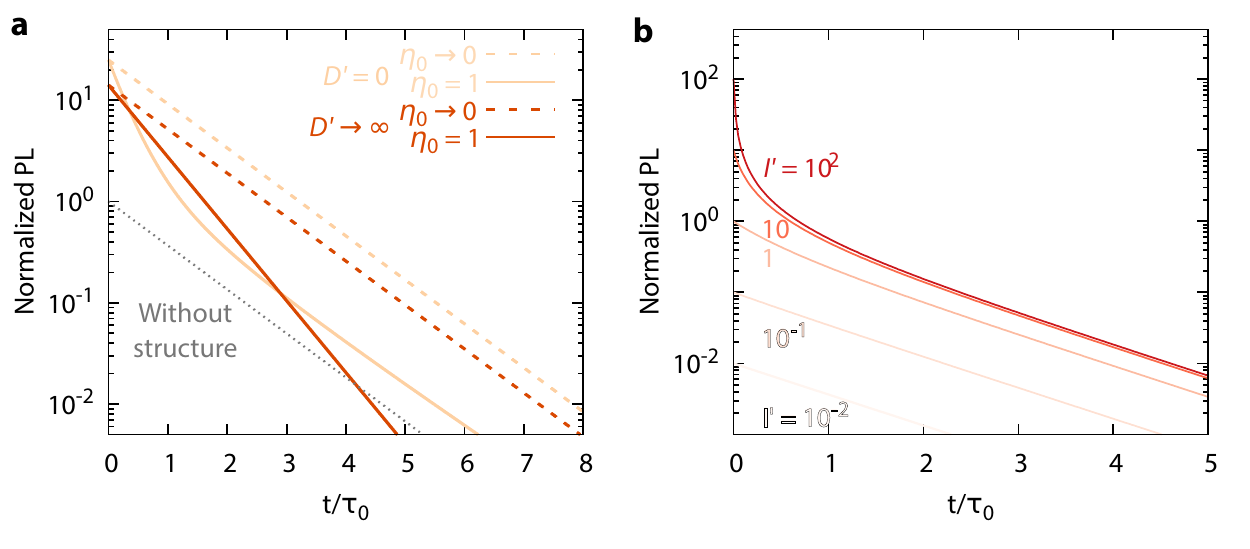}
    \caption{Exciton diffusion and annihilation modify decay rates. (a) Photoluminescence decay traces for an exciton film above the silicon nanodisk array in Figure~2a in the absence of annihilation for limits of quantum yield and diffusion. (b) Photoluminescence decay trace in the presence of annihilation for a bare exciton film with perfect quantum yield as a function of incident power.}
    \label{decaytrace}
\end{figure}

Diffusion affects the modification of decay rate when the film is placed on the nanodisk array (Figure~\ref{decaytrace}a). The bare film in the absence of the nanostructure shows a mono-exponential photoluminescence decay. In the limit of ultra-low quantum yield ($\eta_0 \rightarrow \infty$), both when diffusion is very high or very low, placing the film on the nanostructure modifies only the intensity of photoluminescence but not the decay rate. This is because in this limit, the local decay rate enhancement $F_{decay} = \eta_0 F_{em} + 1 - \eta_0 \rightarrow 1$. As a result, the slope of the photoluminescence decay remains the same on the array. However, for excitons with high quantum yield, the decay rate changes on the array. In the absence of diffusion, the emitters at different locations decay with different rates, resulting in photoluminescence decay which is no longer mono-exponential. In the limit of infinite diffusion, the instant redistribution of exciton density makes the decay mono-exponential again. The increased decay rate results, however, in a different slope compared to the bare exciton film.

Exciton-exciton annihilation modifies the decay rate at high exciton densities even in the absence of nanostructures (Figure~\ref{decaytrace}b). At low incident powers, the photoluminescence decay is mono-exponential because low exciton densities make annihilation negligible. As the incident power increases, annihilation results in significant decay at short time scales.

\section{Decay rate modification in the limit of ultra-low quantum yield due to diffusion}

\begin{figure}[t]
    \centering
    \includegraphics[width=\textwidth]{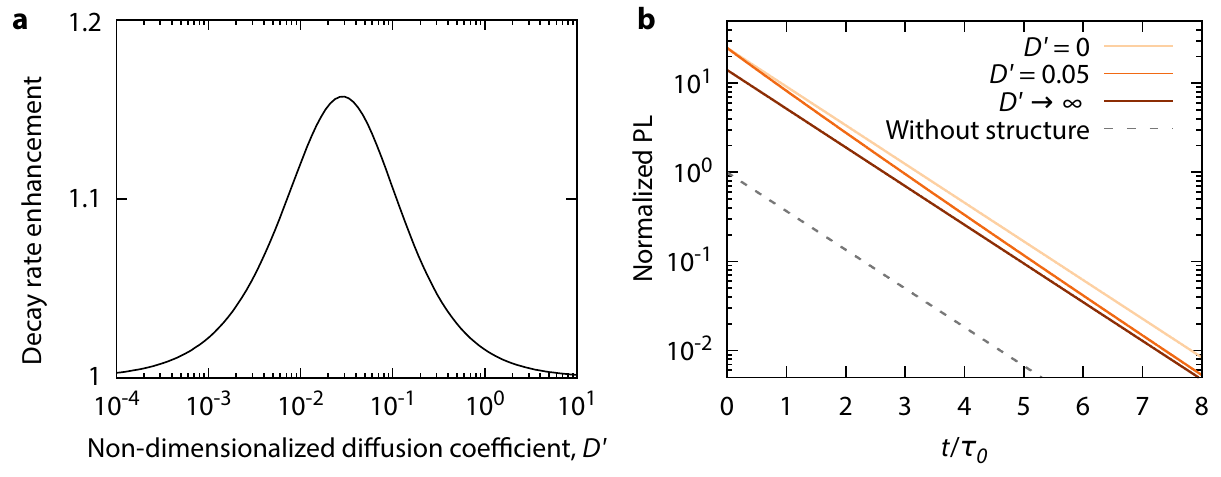}
    \caption{Diffusion modifies the decay rate of excitons near nanostructures even in the limit of ultra-low quantum yield. (a) Decay rate enhancement in the limit of ultra-low quantum yield, and (b) photoluminescence decay for different values of diffusion, for emitters above the array of silicon nanodisks in Figure~3a.}
    \label{zeroetadecay}
\end{figure}

In the presence of diffusion, even the decay rates of emitters with ultra-low quantum yield change near nanostructures (Figure~\ref{zeroetadecay}a). Both in the limit of very low and very high diffusion, there is no decay rate enhancement, as previously seen -- it is only in the intermediate diffusion regime that we observe a modification of decay rate. The reason is that, in the regime of intermediate diffusion, excitons move between regions of different emission enhancements in a finite time, which results in a temporal change in the slope of the photoluminescence decay curve (Figure~\ref{zeroetadecay}b). However, both in the limits of low and high diffusion, the photoluminescence decay is mono-exponential.

\section{Spatially decoupling excitation and emission enhancements using angular dependence of emission}
\begin{figure}[t]
    \centering
    \includegraphics[width=\textwidth]{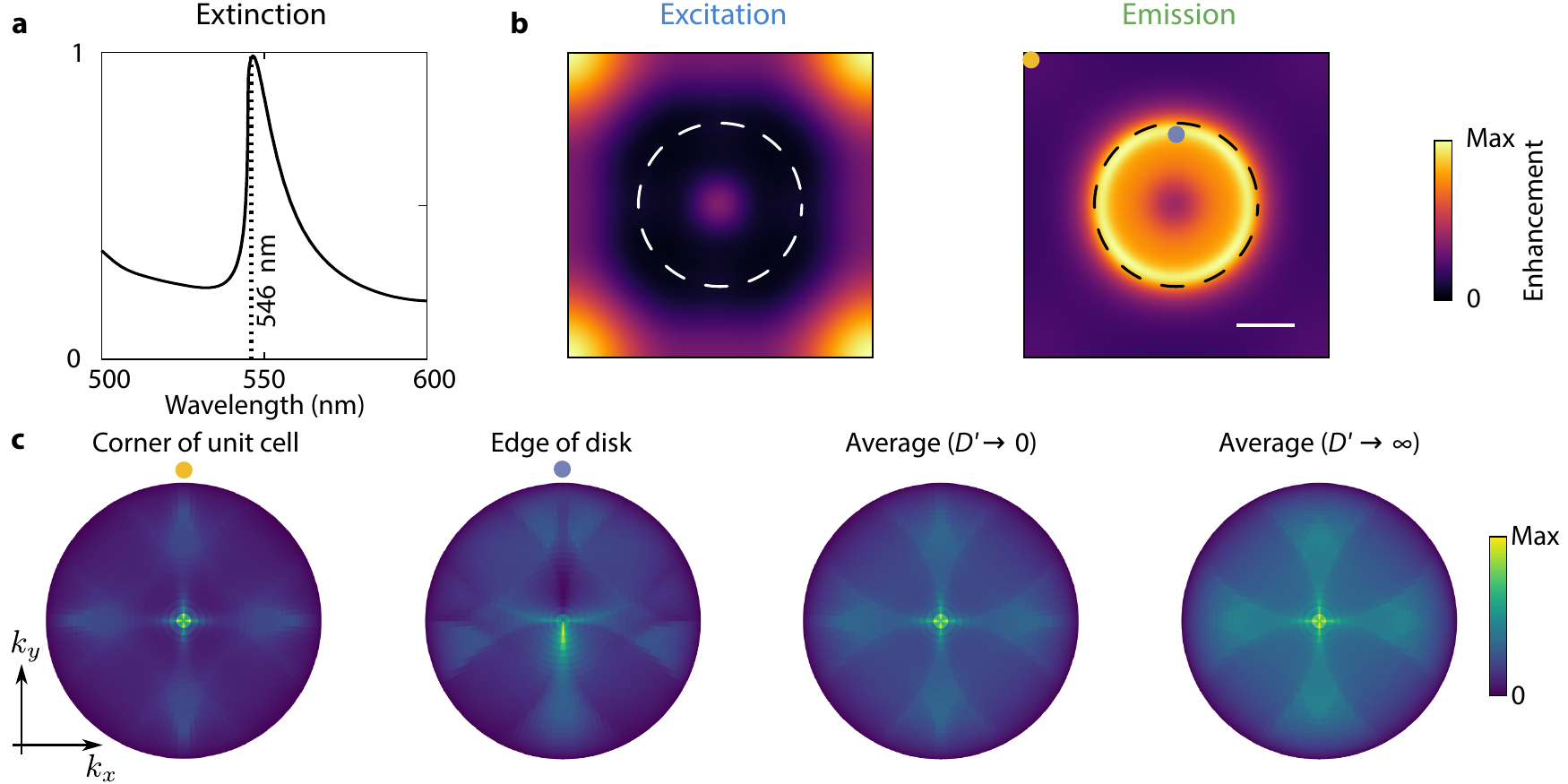}
    \caption{The angular dependence of the emission can be used to spatially decouple excitation and emission enhancements at the same wavelength. (a) Extinction of the array of silicon nanodisks in Figure~3b. (b) The maps of excitation and emission enhancements at $\lambda_{ex} = \lambda_{em} = 546$ nm for emitters above the array. Scale bar 100 nm. (c) Emission patterns in the upper half-space from excitons with ultra-low quantum yield above the array. From left to right: an emitter near the corner of the unit cell (yellow dot), an emitter near the edge of the disk (blue dot), the total emission from the film in the limit of zero diffusion, and in the limit of infinite diffusion.}
    \label{angular}
\end{figure}

The array of silicon nanodisks in Figure~3b shows improved photoluminescence enhancement with diffusion because the excitation and emission enhancements are spatially decoupled (Figure~\ref{angular}b). The excitation and emission wavelengths are equal, $\lambda_{ex} = \lambda_{em} = 546$ nm (Figure~\ref{angular}a). Spatial decoupling of these enhancements occurs in spite of the wavelengths being equal because of the angular variation in the emission from different locations in the unit cell. An exciton near the corner of the unit cell (yellow dot) emits in a narrow angle in the vertical direction (Figure~\ref{angular}c). Meanwhile, an emitter near the edge of the disk (blue dot) emits at an angle slightly away from the vertical. This angle explains why the large emission enhancement near the edge of the disk does not result by reciprocity in an equivalent excitation enhancement under normally incident plane wave illumination from above. In the limit of ultra-low quantum yield and zero diffusion, excitons near the corner of the unit cell dominate the emission. In the limit of infinite diffusion, on the other hand, excitons distribute uniformly in the unit cell before emitting. As a result, more emission away from the vertical occurs in the case of infinite diffusion.

\section{Improving photoluminescence enhancement at high power using emission-dominant enhancement}
\begin{figure}[t]
    \centering
    \includegraphics[width=\textwidth]{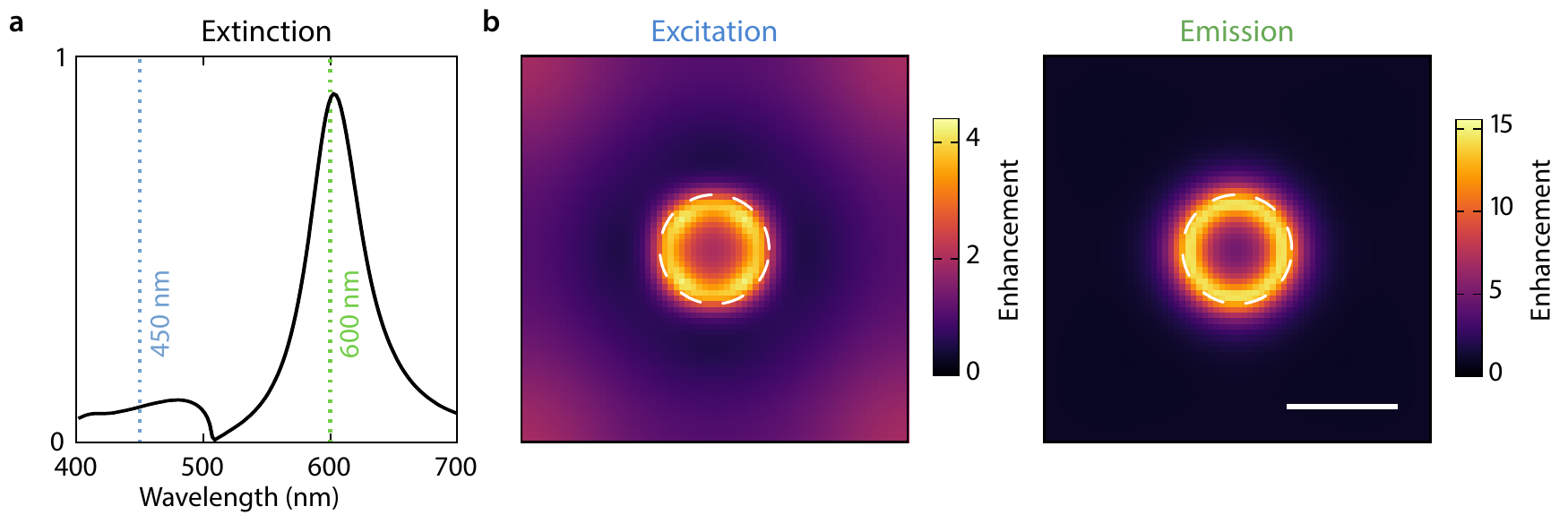}
    \caption{Nanophotonic structures with emission enhancement comparable to or higher than excitation offer improved photoluminescence enhancement with increasing power. (a) Reflectance from, and (b) maps of excitation enhancement at $\lambda_{ex} = 450$ nm and emission enhancement at  $\lambda_{em} = 600$ nm above, for emitters above the array of silver nanodisks in Figure~6b. Scale bar 100 nm.}
    \label{silverenhancement}
\end{figure}

\begin{figure}[b!]
    \centering
    \includegraphics[width=4.75in]{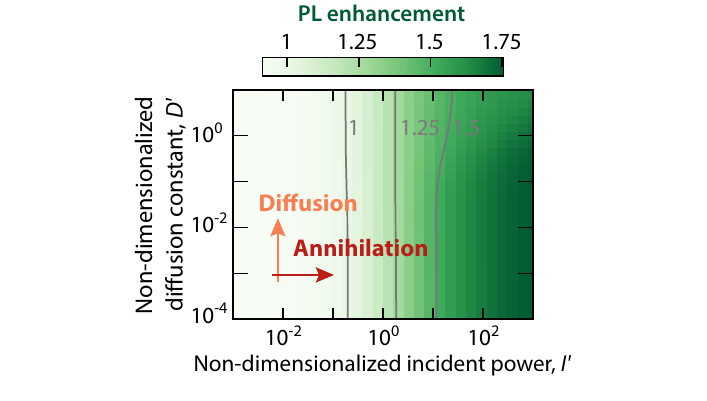}
    \caption{Dielectric nanostructures can also offer improved photoluminescence enhancement with increasing power. Photoluminescence enhancement as a function of incident power under continuous-wave illumination from above for emitters with $\eta_0 = 1$ above an array of silicon nanodisks with $R$~=~110, $H$~=~85, $P$~=~400~nm.}
    \label{silicongood}
\end{figure}

Emitters above the silver nanodisk array in Figure~6b show improved photoluminescence enhancement with increasing incident power. The array has a sharp resonance at the emission wavelength $\lambda_{em}=600$ nm whereas the excitation wavelength  $\lambda_{ex}=450$ nm falls on a different resonance (Figure~\ref{silverenhancement}a). Emission enhancement of the array is higher than the excitation enhancement (Figure~\ref{silverenhancement}b), resulting in the nanophotonic enhancement improving at high powers.

Dielectric nanostructures in which emission enhancement dominates excitation can also show improved photoluminescence enhancement at high incident powers (Figure~\ref{silicongood}).

\section{Nanophotonic enhancement in realistic materials}

\begin{table}[t!]
    \centering
    \bgroup
    \caption{\label{parameters} Representative reported parameters for excitonic materials. Intrinsic quantities are decay rate $\Gamma_0$, quantum yield $\eta_0$, diffusion constant $D$, annihilation constant $\gamma$, and absorption constant $\sigma$. Exciton density $n_0$ and incident power $I_0$ are the characteristic values used to non-dimensionalize the exciton dynamics equation~(1). $D'$ is the non-dimensionalized diffusion constant for period $P=365$ nm. The quantum yield of the MoS$_2$ monolayer on the quartz substrate is unity due to treatment with bis(trifluoromethane)sulfonimide (TFSI)~\cite{Goodman_2020}.}
    \def\arraystretch{1.25} 
    \begin{tabular}{|c|c|c|c|c|c|c|c|c|}
    \hline
    Material & $\Gamma_0$  & $\eta_0$ & $D$  & $\gamma$  & $\sigma$  & $n_0$  & $I_0$  & \multirow{2}{*}{$D'$}\\
    & (ns$^{-1}$) & (\%) & (cm$^2$s$^{-1}$) & (cm$^2$s$^{-1}$) & (J$^{-1}$) & (cm$^{-2}$) & (mW $\mu$m$^{-2}$) & \\
    \hline
    \multicolumn{9}{|c|}{\textbf{Transition metal dichalcogenide monolayers}} \\
    \hline
    WSe$_2$\cite{Mouri_2014} & 0.25 & 0.1 & 2.2 & 0.35 & $2 \times 10^{17}$ & $7.1 \times 10^8$ & $8.9 \times 10^{-6}$ & 6.6\\
    \hline
    WS$_2$\cite{Yuan_2015,Yuan_2017} & 1.24 & 6 & 2 & 0.41 & $9.4 \times 10^{16}$ & $3 \times 10^9$ & $4 \times 10^{-4}$&  1.2\\
    \hline
    MoSe$_2$\cite{Kumar_2014,Kumar_2014b,Roy_2020} & 7.7 & 0.38 & 12 & 0.165 & $5.2 \times 10^{16}$& $4.7 \times 10^{10}$ & $6.8 \times 10^{-2}$ & 1.2\\
    \hline
    MoS$_2$\cite{Goodman_2020} & 0.056 & 100 & 0.06 & 0.8 & $1.8 \times 10^{17}$ & $6.9 \times 10^7$ & $2.1 \times 10^{-7}$ & 0.81\\
    \hline
    \multicolumn{9}{|c|}{\textbf{Two-dimensional perovskites}\cite{Deng_2020}} \\
    \hline
    1 layer & 0.45 & 0.79 & 0.06 & $1.7 \times 10^{-2}$ & $2.6 \times 10^{16}$ & $2.6 \times 10^{10}$ & $4.5 \times 10^{-3}$ & 0.10 \\
    \hline
    2 layers & 0.23 & 0.79 & 0.07 & $4.1 \times 10^{-3}$ & $3.2 \times 10^{16}$ & $5.6 \times 10^{10}$ & $4.0 \times 10^{-3}$ & 0.23 \\
    \hline
    3 layers & 0.3 & 0.79 & 0.15 & $3.5 \times 10^{-3}$ & $3.5 \times 10^{16}$ & $8.6 \times 10^{10}$ & $7.3 \times 10^{-3}$ & 0.38 \\
    \hline
    4 layers & 0.2 & 0.79 & 0.25 & $2.8 \times 10^{-3}$ & $3.1 \times 10^{16}$ & $7.1 \times 10^{10}$ & $4.6 \times 10^{-3}$ & 0.94 \\
    \hline
    5 layers & 0.15 & 0.79 & 0.34 & $1.3 \times 10^{-3}$ & $3.2 \times 10^{16}$ & $1.2 \times 10^{11}$ & $5.3 \times 10^{-3}$ & 1.7 \\
    \hline
    \end{tabular}
    \egroup
\end{table}

Diffusion and annihilation strongly modify the photoluminescence enhancement of realistic excitonic materials in nanostructured landscapes. We tabulate the intrinsic excitonic parameters of some transition metal dichalcogenide monolayers and two-dimensional perovskites from various references in Table~\ref{parameters}. As the sources do not provide the absorption constant $\sigma$ directly, we calculate it from the absorptance by assuming that every absorbed photon generates an exciton~\cite{Yuan_2015}. Along with the intrinsic parameters, we also show the characteristic exciton density $n_0$ and the characteristic power $I_0$ used for non-dimensionalization, as well as the non-dimensionalized diffusion constant $D'$ for $P=365$ nm.

\begin{figure}[t!]
    \centering
    \includegraphics[width=4.25in]{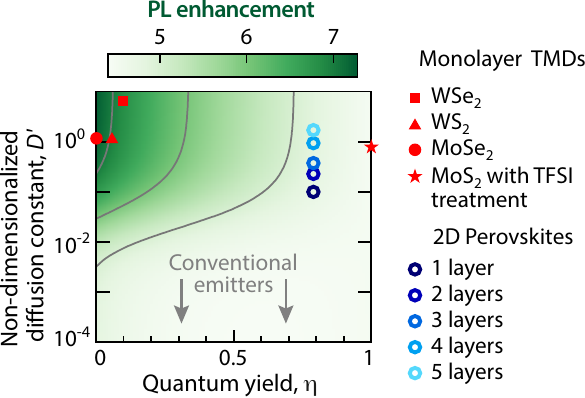}
    \caption{Realistic excitonic materials are highly diffusive, which modifies their photoluminescence enhancement. Excitonic parameters for TMD monolayers and 2D perovskites, overlaid on the photoluminescence enhancement of the array of silicon nanodisks (Figure~4c).}
    \label{realdiffusion}
\end{figure}

The non-dimensionalized diffusion constants for the all the TMD monolayers are of the order of unity (Table~\ref{parameters}). As a result, the silicon array in Figure~4c optimized for highly diffusive excitons provides strong photoluminescence enhancement for TMD monolayers with low quantum yield (Figure~\ref{realdiffusion}). Treating the excitons as conventional immobile emitters would significantly underestimate their photoluminescence enhancement in this case. Emitters with high quantum yield such as 2D perovskites and TFSI-treated MoS$_2$ do not see a significant modification of photoluminescence enhancement on account of their high diffusion.

\begin{figure}[t!]
    \centering
    \includegraphics[width=4.25in]{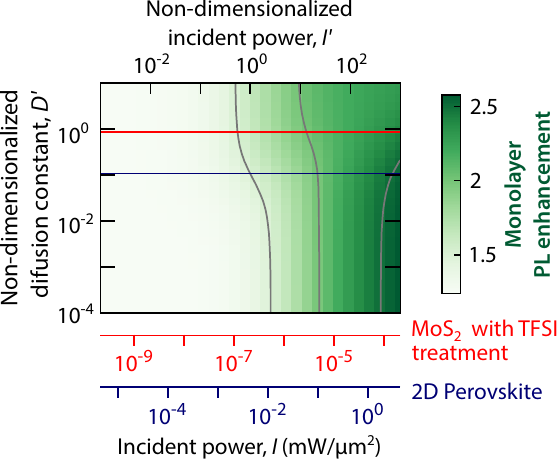}
    \caption{Realistic excitonic materials with high quantum yield show annihilation at typical illumination powers. Photoluminescence enhancement of the array of silver nanodisks (Figure~6b), overlaid with the illumination powers ($I$) of monolayers of MoS$_2$ and 2D perovskites, corresponding to the non-dimensionalized incident power $I'$.}
    \label{realannihilation}
\end{figure}

Typical excitonic emitters with high quantum yield also suffer from strong exciton-exciton annihilation (Table~\ref{parameters}). Due to the low intrinsic decay rate and high annihilation in MoS$_2$ monolayers treated with TFSI, annihilation effects become significant at very low powers of the order of a few nW/$\mu$m$^2$ (Figure~\ref{realannihilation}). At typical powers of the order of $\mu$W/$\mu$m$^2$, the silver nanodisk array in Figure~6b designed to be emission dominant shows high photoluminescence enhancement. Monolayers of 2D perovskites show much lower annihilation, which nevertheless becomes significant at high powers of the order of mW/$\mu$m$^2$. The interplay between diffusion and annihilation also becomes relevant at intermediate powers for both materials due to their high non-dimensionalized diffusion constants.

\section{Dimensionality of emitter orientations}

As excitons in typical materials do not have a preferential orientation, we compute the average excitation enhancement in the main text from the total electric field intensity (see Methods). Similarly, we computed the emission enhancement by averaging emitters along all possible orientations in three dimensions. However, the direct transitions in TMD monolayers occur through in-plane electric fields only. As a result, their excitation enhancement only depends on the in-plane electric field intensity~\cite{Raziman_2019} and their emission enhancement will be the average enhancement of dipoles along all two-dimensional orientations.

\begin{figure}[t!]
    \centering
    \includegraphics[width=4.25in]{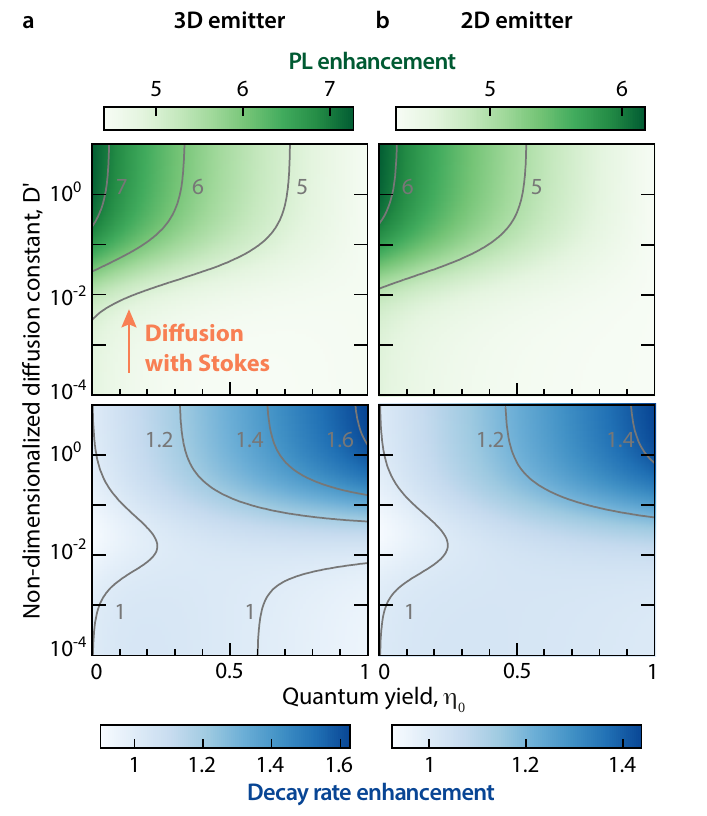}
    \caption{Nanophotonic enhancements of dipolar emitters orientationally averaged in two dimensions have a similar qualitative dependence with exciton dynamics as emitters orientationally averaged in three dimensions. Enhancements of photoluminescence and total decay rate under excitation at $\lambda_{ex}$ = 461 nm and emission at $\lambda_{em}$ = 553 nm for the silicon nanodisk array in Figure~3a for emitters orientationally averaged in (a) three dimensions (corresponding to Figure~4c), and (b) in two dimensions.}
    \label{compare3d2d}
\end{figure}

Although the dimensionality changes the numerical value of enhancements, the qualitative behavior remains the same (Figure~\ref{compare3d2d}). Physical effects we discussed in the main text such as diffusion decoupling excitation and emission enhancements and annihilation suppressing the relevance of excitation are independent of the whether we consider two-dimensional or three-dimensional emitters.

\section{Emission enhancement counteracts diffusion}

In systems with high emission enhancement, the reduced lifetime of the exciton lowers the effective diffusion length as well. As a result, increasing the emission enhancement can turn an emitter in the $D'\rightarrow \infty$ limit to one in the $D'\rightarrow 0$ limit.

\begin{figure}[t!]
    \centering
    \includegraphics[width=4.25in]{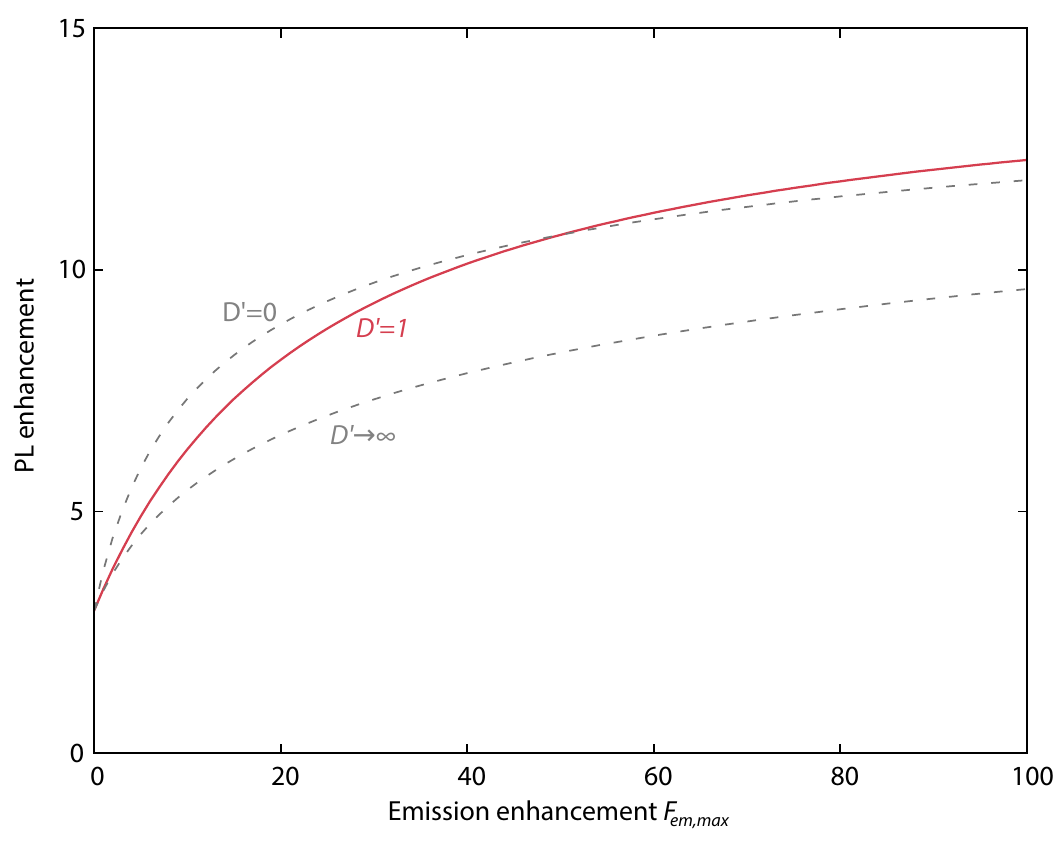}
    \caption{Increasing the emission enhancement counteracts the effect of diffusion. For an excitonic material with $D'=1$ and $\eta_0=0.2$, increasing the Purcell factor takes PL enhancement from the infinite diffusion limit to the zero diffusion limit.}
    \label{Purcellincrease}
\end{figure}

As an example, consider a system where excitation and emission enhancements are both Gaussians at the centre of the unit cell with half-widths equal to 25\% of the periodicity. We take the exciton parameters $D'=1$ and $\eta_0=0.2$. Keeping the height $F_{ex,max}$ of the excitation enhancement Gaussian fixed at 10, we vary the height $F_{em,max}$ of the emission enhancement Gaussian. At low values of emission enhancement, excitons can diffuse throughout the unit cell, making the PL enhancement similar to the case of infinite diffusion (Figure~\ref{Purcellincrease}). However, when the emission enhancement increases, the excitons decay quicker, reducing the distance they can diffuse. As a result, the PL enhancement becomes similar to the limit of zero diffusion.

\section{Photoluminescence enhancement in nanowires}
\begin{figure}[t!]
    \centering
    \includegraphics[width=4.25in]{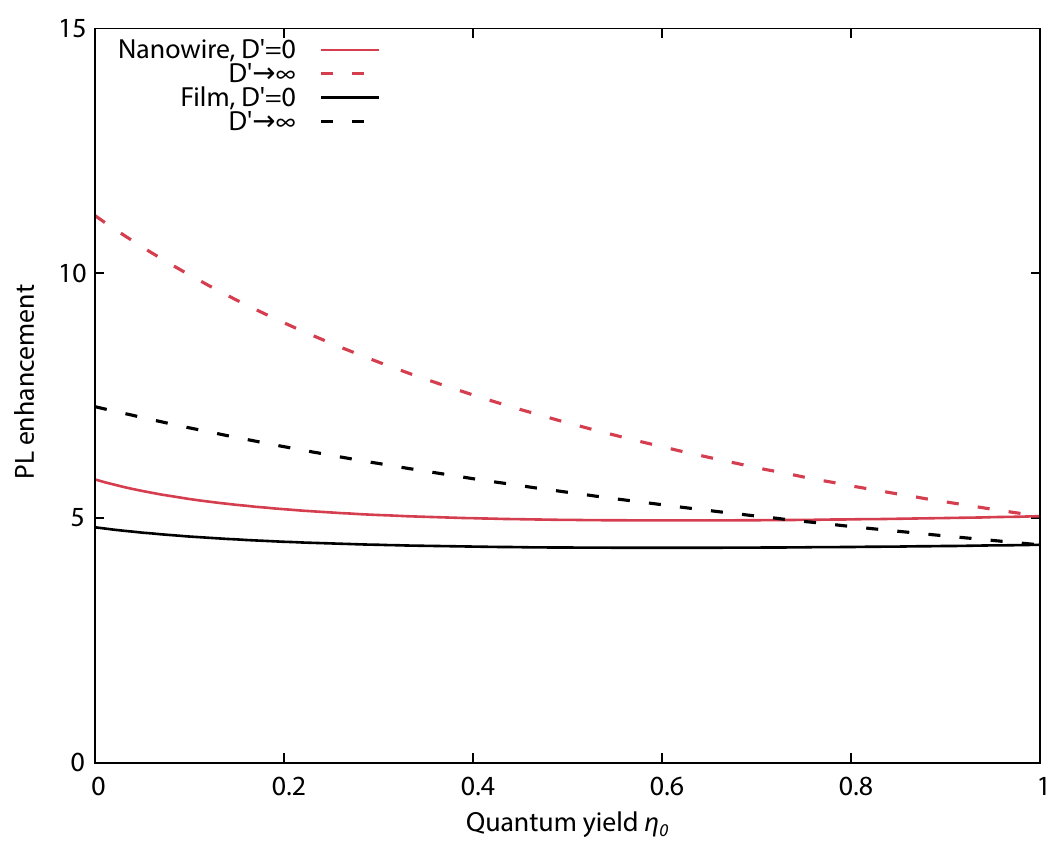}
    \caption{When the film of excitonic material above the silicon nanodisk array in Figure~3a is replaced by a long nanowire along the diagonal of the unit cell, the dependence of PL enhancement ($\lambda_{ex}$ = 461 nm, $\lambda_{em}$ = 553 nm) on diffusion remains the same qualitatively.}
    \label{wire}
\end{figure}

Although we have discussed the interplay of exciton dynamics and nanophotonic enhancement in the context of excitonic films, the underlying principles can be extended to other geometries as well. Nanowires are ubiquitous one-dimensional excitonic structures. We consider a long nanowire along the diagonal of the silicon nanodisk array in Figure~3a, spanning many unit cells so that we can treat it as infinitely long. We neglect scattering effects at the two ends of the nanowire and compute the PL enhancements at the zero and infinite diffusion limits for $\lambda_{ex}$ = 461 nm, $\lambda_{em}$ = 553 nm (corresponding to Figure~4c). As the average excitation and emission enhancements are higher on the diagonal compared to other regions (Figure~4b), PL enhancement is higher for the nanowire than the film (Figure~\ref{wire}). Qualitatively, both the film and the  wire show similar behavior of PL enhancement increasing with diffusion and decreasing with quantum yield.

\bibliographystyle{nature}
\bibliography{references}